\documentclass[prl,superscriptaddress,showpacs,twocolumn,showkeys,10pt]{revtex4-1}
\usepackage{amsfonts,amsmath,amssymb,graphicx,verbatim,epstopdf}
\usepackage[english]{babel}

\begin{document}

\title{Fermi edge polaritons in a highly degenerate 2D electron gas: a diagrammatic theory}
\author{Maarten Baeten}
\affiliation{TQC, Universiteit Antwerpen, Universiteitsplein 1, B-2610 Antwerpen, Belgium}
\author{Michiel Wouters}
\affiliation{TQC, Universiteit Antwerpen, Universiteitsplein 1, B-2610 Antwerpen, Belgium}
\date{\today}

\begin{abstract}
We present a theoretical study on polaritons in highly doped semiconductor microcavities. In particular, we focus on a cavity mode that is resonant with the absorption threshold (`Fermi edge'). In agreement with experimental results, the strong light-matter coupling is maintained under very high doping within our ladder diagram approximation. While the lower polariton is qualitatively unaltered, it acquires a finite lifetime due to relaxation of the valence band hole if the electron density exceeds a certain critical value. On the other hand the upper polariton has a finite lifetime for all densities, because it lies in the electron-hole continuum where no bound state exists. Our calculations show that a narrow upper polariton quasiparticle still exists as a result from the interplay between light-matter coupling and final state Coulomb interaction.
\end{abstract}

\maketitle

Microcavity polaritons are bosonic quasiparticles that appear in a semiconductor microcavity with an embedded quantum well (QW) when the cavity is tuned to an excitonic transition in the QW. In the last decade they have emerged as quasiparticles with many favorable properties for studying quantum optics and many-body physics in integrated photonic structures.
Bose-Einstein condensation of polaritons and their superfluid properties have been investigated in detail both experimentally and theoretically \cite{iac_review}.
In these studies, the tunability of the polariton properties is often exploited. For example, the number of quantum wells is varied to change the Rabi frequency; etching, strain and surface acoustic waves have been used to create potentials for the polaritons. 

A tuning parameter that has received relatively little attention is the introduction of charges in the QW that interact with the polaritons. In the case of low charge concentration, trionic bound states exist and a trionic polariton is observed \cite{Rapaport,Bloch}. An experiment with high density modulation doping was performed by Gabbay {\em et al.} \cite{GabbayCohen}. They have shown that strong light-matter coupling is possible, even though neither an excitonic nor a trionic state could be resolved. 

From the theoretical side, the mixed electronic-polaritonic system was proposed to reach superconductivity at higher temperatures, possibly even room temperature, thanks to a strongly attractive electron-electron interaction mediated by the polaritons \cite{laussy}. For what concerns the effect of the electron gas on the polaritons, it has been suggested to facilitate polariton lasing thanks to the polariton-electron scattering \cite{Kavokin}. In this work, the polariton quasiparticle itself was assumed not to be affected by the electron gas. A first theoretical study on the modification of the single particle polariton properties due to the electron gas started from an estimate of the oscillator strength computed in the context of the Mahan singularity \cite{glazov,Averkiev}. Indeed, the optical excitation of an electron gas (metal or doped semiconductor) is a central problem in many-body physics that was introduced by Mahan \cite{mahan} and to which Nozi\`eres, De Dominicis \cite{nozieres} and Anderson \cite{Anderson} made seminal contributions. It is still an active topic of research that has recently attracted the attention of the community working on ultracold atomic gases \cite{knap,demler,Schirotzek}.

\begin{figure}[htbp]
\centering
\includegraphics[scale=0.25]{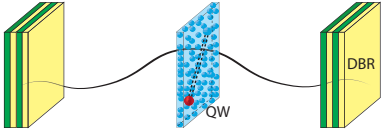}
\caption{Microcavity containing a highly n-doped QW. The system oscillates coherently between the state with a photon in the microcavity and the state with a valence band hole and an additional electron in the QW.}
\label{fig:MC}
\end{figure}

The purpose of this Letter is to study the different regimes of light-matter coupling in highly doped semiconductor QWs, see Fig. \ref{fig:MC}. The regime of low doping density, where trionic correlations are dominant, is not in the scope of this work.
A first effect of the presence of the high density electron gas is that the low-lying momentum states cannot be used for the formation of the exciton, because of the Pauli principle. More importantly, the electron gas has, in contrast to the empty QW, low lying particle-hole excitations. They are responsible for the screening of the electron-hole Coulomb interaction. Both the Pauli blocking and the screening reduce the exciton binding energy \cite{kleinman,kulakovskii}. 
As long as the particle-hole excitations are virtual, they do not qualitatively alter the light-matter coupling, but merely renormalize the Rabi frequency. Qualitative changes occur when the polariton energy becomes sufficiently high to excite free electron-hole pairs. 

\begin{figure}[h!]
\centering
\includegraphics[height=45mm,width=0.9\columnwidth]{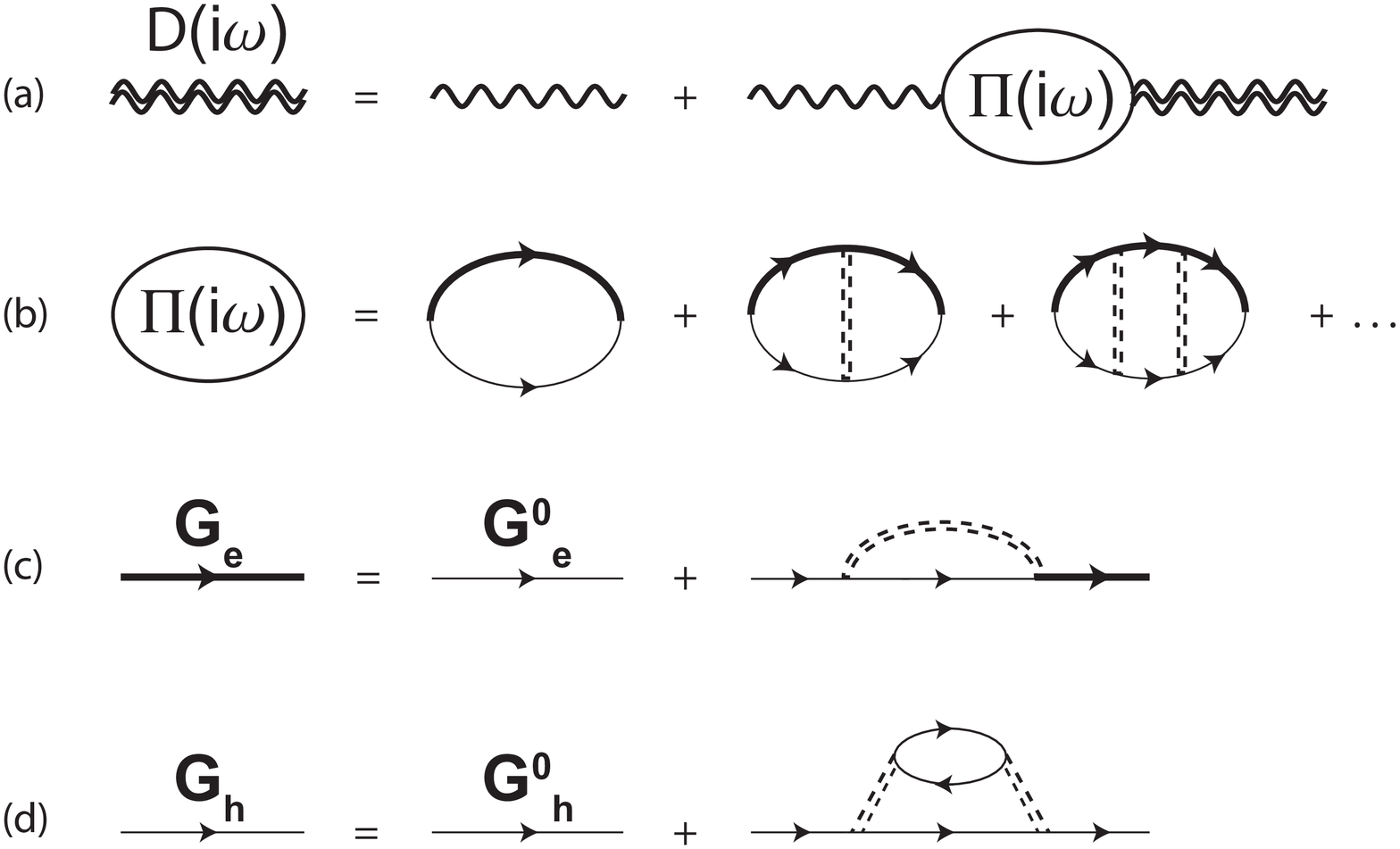}
\caption{(a) Dyson series for the photon propagator $D\left(i\omega\right)$. (b) The photon self energy $\Pi(i\omega)$ is obtained in the ladder diagram approximation. (c,d) Electron $G_e$ and hole $G_h$ Green's functions used in the calculations. }
\label{Fdiag}
\end{figure}

A systematic approach to study the different regimes is diagrammatic perturbation theory \cite{mahanBook}. In this formalism, excitonic effects are taken into account by including the electron-hole Coulomb interaction through ladder diagrams (see Fig. \ref{Fdiag}). The screening of the Coulomb interaction can be taken into account by a dielectric function, that we approximate by the RPA result for the jellium model. Furthermore, we assume that the quantum well electrons and holes are purely two-dimensional. To be specific, we will consider n-doping, so that only a free electron gas is present and a single photo-created hole. 
The primary physical quantities of interest, the absorption and photoluminescence spectrum, can be characterized through the photon spectral function, defined as
\begin{equation}
A(\omega) = -\frac{1}{\pi}\,\text{Im}D^{ret}(\omega),
\label{SF}
\end{equation}
with $D^{ret}(\omega)=D(i\omega\to\omega+i\eta^+)$ the retarded photon propagator, given by
\begin{equation}
D\left(i\omega\right) = \frac{1}{i\omega-\omega_c - \Pi(i\omega)}. 
\label{PhotonProp}
\end{equation}
Here $\omega_c$ corresponds to the cavity mode which is assumed to be lossless. The photon momentum is taken to be zero for simplicity.
The photon self energy $\Pi(i\omega)$ originates from the material excitations within the quantum well and thus contains all many-body physics of the electron gas. We will approximate it by the Feynman diagrams shown in Fig. \ref{Fdiag}.
The Coulomb interaction in these diagrams is already taken to be the RPA screened interaction, indicated by the double dashed lines. 
While for the electron a complete Dyson series for the screened exchange interaction is taken into account, for the hole both the Dyson series for the propagator and the dynamical self energy are expanded to lowest order to keep the calculations manageable:
\begin{eqnarray}
G_h({\bf{k}},i\nu_n)  &= &  G^0_h({\bf{k}},i\nu_n)\crcr
&+& G^0_h({\bf{k}},i\nu_n)\,\Sigma^{RPA}({\bf{k}},i\nu_n)\,G^0_h({\bf{k}},i\nu_n) \label{eq:Gh}
\end{eqnarray}
\vspace{-1cm} 
\begin{eqnarray}
\Sigma^{RPA}({\bf{k}},i\nu_n) &=&  \Sigma_{exchange}({\bf{k}})+\Sigma_{1 bubble}({\bf{k}},i\nu_n).
\label{eq:SigmaRPA}
\end{eqnarray}
Obviously, this expansion is only valid for small self energy. The approximation \eqref{eq:Gh},\,\eqref{eq:SigmaRPA} can be shown to be equivalent to the use of Fermi's golden rule to calculate the linewidth of a polariton due to the relaxation of the hole by the creation of one particle-hole excitation in the electron Fermi sea (see supplemental material).
The dynamical self energy of the electron is omitted, because its relaxation is suppressed by Pauli blocking. 

Under the assumption that the final state Coulomb interaction (FSI) is frequency-independent, a Bethe-Salpeter equation (BSE) for the electron-hole propagator can be derived \cite{mahan}. It reads
\begin{equation}
[\omega - H_X({\bf{k}},{\bf{k^\prime}})]P({\bf{k}},i\omega) = g,
\label{eq:P}
\end{equation}
with
\begin{eqnarray}
H_X &=& \frac{\hbar^2k^2}{2m_r} + \Sigma_x({\bf{k}}) +V^s_{C}\left(\bf{k,k^\prime}\right)+ i\Sigma_h({\bf{k}},\omega)
\label{eq:HX}
\end{eqnarray}
and
\begin{widetext}
\begin{eqnarray}
\Sigma_h({\bf{k}},\omega)\propto 
\sum_{{\bf{q,K}}}(V^s_{\bf{q}})^2 \frac{n_{F}(\varepsilon^e_{\bf{K+q}})[1-n_{F}(\varepsilon^e_{\bf{K}})]n_{F}(-\varepsilon_{\bf{-k+q}}^{h}+\varepsilon^e_{\bf{K+q}}-\varepsilon^e_{\bf{K}})}{(\varepsilon_{\bf{-k+q}}^{h}-\varepsilon^e_{\bf{K+q}}+\varepsilon^e_{\bf{K}}-\varepsilon_{\bf{-k}}^{h})^2}\,\delta[\omega-\varepsilon_{\bf{-k+q}}^{h}+\varepsilon^e_{\bf{K+q}}-\varepsilon^e_{\bf{K}}-\varepsilon^e_{\bf{k}}-\Sigma_{x}(\bf{k})].
\label{eq:sigmahole}
\end{eqnarray}
\end{widetext}
Here, the Hamiltonian $H_X$ \eqref{eq:HX} describes the relative part of the two-particle system; the reduced mass is $m_r=1/(m_e^{-1}+m_h^{-1})$. 

To arrive at the BSE, we had to make the static approximation for the FSI, for which we used the 2D RPA expression of Stern \cite{Stern}. This static approximation simplifies the subsequent calculations, but it will be interesting to explore the physical consequences of relaxing this approximation. A well known limitation of RPA is that it is only valid for a high density electron gas \,$a^*_0k_F>>1$ with $a^*_0$ being the effective Bohr radius. Specifically for this problem, it means that RPA is not able to capture trionic effects that become important at low density.

The photon self energy is finally related to the vertex factor $P({\bf{k}},i\omega)$ as
\begin{eqnarray}
\Pi(i\omega) &=& \frac{g}{S}\sum_{\bf{k}} P({\bf{k}},i\omega).
\label{Polarization}
\end{eqnarray}
Here $g$ is the light-matter coupling and $S$ the quantum well surface. 

\begin{figure}[htbp]
\centering
   \includegraphics[height=50mm,width=1\columnwidth] {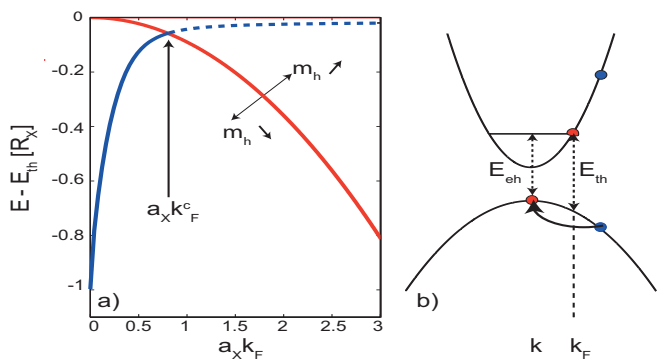}
 \caption{Below $a_Xk_F^c$ the exciton forms the groundstate of the optically excited quantum well. Above the critical wavevector the exciton becomes metastable (dashed line) and obtains a finite lifetime due to relaxation of the hole via emission of a particle-hole pair (right figure). $a_Xk_F^c$ is determined by $m_e/m_h$. In the figure $m_e/m_h=0.13$ is used. 
 }
 \label{fig:evskf}
\end{figure}

It can be seen from Eq. (\ref{eq:P}) that the bound excitonic levels show up as poles of $P({\bf{k}},i\omega)$ and consequently as poles in the photon self energy $\Pi(i\omega)$. For an undoped QW this can be calculated analytically \cite{Citrin}; here we implement it numerically. The lowest lying exciton level is depicted in Fig. \ref{fig:evskf}a as a function of $a_Xk_F$ (blue line), where $a_X/a_0^*=(1+m_{e}/m_{h})/2$ holds. Its energy is measured in exciton Rydbergs $R_X=\hbar^2/2m_r a_X^2$.

A first important energy scale in the exciton physics is the lowest energy of a free electron-hole pair, indicated as $E_{\textrm {eh}}$ in Fig. \ref{fig:evskf}b and by a red line in panel a. 
To be consistent with the computation of the bound exciton, we have included the renormalization of the Fermi energy due to the exchange interaction. The zero of energy in Fig. \ref{fig:evskf}a was chosen as the energy of a free electron-hole pair with both particles at the Fermi wave vector. This energy corresponds to the absorption threshold $E_{\text{th}}$ (`Fermi edge') in Fig. \ref{fig:evskf}b. 

Below a critical density, the electron and hole form a stable bound state below $E_{\textrm {eh}}$, the well-known exciton. In this case, the hole self energy (\ref{eq:sigmahole}) vanishes at the bound state energy, corresponding to an infinite exciton life time.
Beyond the critical value $a_Xk_F^{c}$, which depends on the ratio $m_e/m_h$ (trend indicated with a black arrow), the exciton crosses the free electron-hole energy and is no longer the lowest energy state, but becomes a metastable bound state (blue dashed line). It is the hole kinetic energy that is responsible for the fact that the exciton is not the ground state. When the hole is bound to the electron in a zero momentum exciton, the momentum of the electron has to be compensated by an opposite momentum of the hole. Because the low momentum electron states are Pauli blocked, there is a minimal kinetic energy cost $k_F^2/2m_h$ for the hole to form a bound state with the electron. Obviously, this cost vanishes in the limit where the hole becomes infinitely heavy $m_h\to\infty$. We then find that the exciton is stable for arbitrary density and corresponds to the bound state that always exists for an attractive potential in 2D.

When the exciton is metastable, the hole self energy (\ref{eq:sigmahole}) is nonzero at the exciton energy. It is seen to correspond to a relaxation process through the emission of a low energetic particle-hole pair, see Fig. \ref{fig:evskf}b. Note that only the lifetime contribution of this process is taken into account and the associated energy shift has not been considered in this work. The delta function in \eqref{eq:sigmahole} expresses energy and momentum conservation for this process, while the occupation factors describe the amount of phase space that is available for the decay process. Below the critical Fermi wave vector $k_F^c$, this phase space factor vanishes when energy and momentum are conserved. The relaxation of the hole by plasmon emission was considered in the plasmon-pole approximation \cite{Overhauser}. However, this relaxation channel does not contribute because it does not conserve energy and momentum.
\begin{figure*}[htb!]
\centering
   \includegraphics[height=55mm,width=2.15\columnwidth] {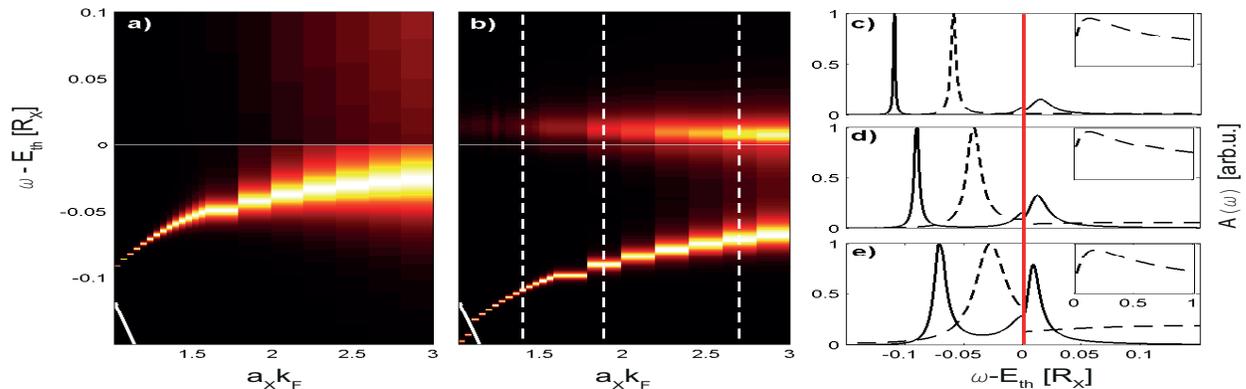}
 \caption{a) Photon spectral function as a function of electron Fermi wave vector and energy (arb. units). The free electron-hole energy $E_{\textrm{eh}}$ is indicated with the solid white line in the left lower corner. b) shows the photon spectral function at $\omega_c=E_{\textrm {th}}$. The lower and upper `Fermi edge polaritons' are clearly visible. Different cuts along $a_Xk_F$ (white dashed lines) show the narrowing of the photon spectral function (black solid) under the continuum (red line) with respect to the exciton spectral function (black dashed). The excitonic enhancement is shown in more detail in the insets. Note the different scales showing that the upper polariton has a much narrower lineshape as compared to the excitonic enhancement.
}
 \label{fig:MainFigure}
\end{figure*}
Unfortunately, the hole self energy (\ref{eq:sigmahole}) diverges for all energies higher than $E_{\textrm {th}}$ (`the continuum') so that we cannot use our previous approach to include the hole self energy. A more sophisticated approach is necessary. As a first approximation, we follow Schmitt-Rink \cite{SchmittRink} and obtain a first meaningful result by omitting the hole self energy altogether.
Actually above $E_{\textrm{th}}$, the photon self energy acquires an imaginary part even when the Hamiltonian (\ref{eq:HX}) is real. Physically, this finite lifetime comes from the conversion of photons into free electrons and holes.  Therefore, as a lowest order approximation to the spectral function in this regime we calculate the spectral function exactly in the same way as before, summing up all ladder diagrams to infinite order, but the self energy for the hole is now discarded. The solution of the BSE equation can now be calculated analytically to some extent \cite{SchmittRink} but we implemented it straightforwardly numerically. For the exciton, we recover the `excitonic enhancement' \cite{SchmittRink,Hawrylak}, an increase in the absorption close to $E_{\textrm{th}}$. This is a many-body modification, due to final state interactions, to the constant absorption obtained in a single particle picture. 

A summary of the exciton physics is presented in Fig. \ref{fig:MainFigure}a. It shows the photon spectral function as a function of energy and the electron Fermi wave vector, in the limit of very large positive cavity detuning. In this limit, we recover the exciton physics. The exciton is blue shifted when the electron density increases. It is also clear that the excitonic linewidth then increases. The excitonic enhancement for $\omega>E_{\text{th}}$ is not very clear on this figure and is shown in more detail in the inset on Fig. \ref{fig:MainFigure}c-e. It is an intriguing question whether this excitonic effect may lead to the formation of polaritons. Experimentally, this question was answered positively in Ref. \cite{GabbayCohen} where the observation of `Fermi edge polaritons' was claimed.

To answer this question theoretically, we plot in Fig. \ref{fig:MainFigure}b the photon spectral function when the cavity photon is in resonance with the Fermi edge $E_{\text{th}}$.  Different detunings are shown in the supplemental material. Several cuts at fixed electron densities, indicated with white dashed lines, are shown in the panels (c-e). As expected, the lower polariton is pushed below the excitonic resonance. The shift of the polariton with respect to the exciton is however reduced due to the presence of the electron gas. This is expected since the light-matter coupling is proportional to the electron-hole overlap, which is reduced due to Pauli blocking and screening. Importantly, the linewidth of the polariton is strongly reduced with respect to the excitonic linewidth. This narrowing can be attributed to the reduced final state phase space that slows the hole relaxation. For low electron gas densities it is even possible for the polaritons to decrease their energy below the free electron-hole energy $E_{\text {eh}}$ making their linewidth vanish (on the figure a very small non-zero value is added to make the resonance visible). The linewidth reduction can be seen more clearly from the various cuts taken for different values of $a_Xk_F$ (Fig. \ref{fig:MainFigure}c-e). Qualitatively however the lower polariton is entirely similar to the one in the empty quantum well.

This is not the case for the upper polariton that can be clearly identified in Fig. \ref{fig:MainFigure}b, above the absorption threshold. Above $E_{\textrm{th}}$ bound states do no longer exist and one should think of the excitonic enhancement as the valence band hole interacting with all electrons in the Fermi sea {\sl{simultaneously}}. Nevertheless, this polaritonic resonance has a linewidth that is much smaller than the excitonic enhancement feature (see insets of Fig. \ref{fig:MainFigure}c-e). Moreover, it is shifted from the cavity resonance by $\Delta E \approx 0.01$ $R_X$ . Despite the two different approximations that were used to compute the spectral function below and above the absorption continuum (red vertical line) respectively, the matching of the curves is satisfactory. 

It is conceptually interesting to elucidate the role of the final state interactions on this upper polariton. To this purpose, we have turned off the final state Coulomb interaction [$V_C^s=0$ in Eq. (\ref{eq:HX})] in our calculations. Without FSI
a much smaller energy shift from the cavity resonance is observed and a larger linewidth is obtained. It is really the final state Coulomb interactions between the photo-excited electron-hole pair that pushes the quasi-particle above the cavity frequency. We have also checked that the `Rabi splitting' vanishes when the cavity energy is blue detuned with respect to $E_{\text{th}}$. It is indeed the Fermi edge $E_{\textrm{th}}$  that is responsible for the formation of an upper polariton in the continuum, see supplemental material. 
Finally, it is worth mentioning that when the light-matter coupling is increased, the upper polariton is further blue shifted and its linewidth increases. For a good visibility of the upper `Fermi edge' polariton, it is actually advantageous if the Rabi splitting is not too large (not too many quantum wells). This trend is opposite to the one of the lower polariton, where a larger Rabi splitting leads to a lower energy and therefore a longer lifetime.

Note that the upper polariton energy decreases in Fig. \ref{fig:MainFigure}b for increasing $k_F$, even though the photon energy is constant and the exciton energy increases. In an undoped quantum well, the opposite dependence is observed for increasing exciton energy. We attribute the decrease of the upper polariton energy to the reduction of the electron-hole overlap when the density is increased.

Our theoretical results are in qualitative agreement with the experiments in Ref. \cite{GabbayCohen}, where at high doping density both an upper and lower polariton were observed. In this experimental work, a phenomenological explanation based on the Fermi edge enhancement of the optical response was suggested. We have shown here that this enhancement is due to the usual excitonic bound state for the lower polariton, where it is of a different, collective nature for the upper polariton.

The discontinuity in the spectral function at $E_{\textrm{th}}$ is of course an artifact of our different approximations below and above this continuum energy. It should be cured by a consistent inclusion of the self energy below and above the absorption threshold energy. This self energy should actually contain the diagrams that are relevant for the physics of the Mahan singularity so to capture the effect of this physics on the nature of the polaritons in doped quantum wells. The present work is only a first step to such a complete description. It will also be of great interest to investigate the interactions between polaritons in doped quantum wells. It may be expected that interactions that are induced by excitations in the electron gas may enhance the polariton-polariton interactions with possible applications in quantum information processing.


\onecolumngrid
\newpage

\begin{center}
{\huge\bf {Supplemental material}}
\end{center}

\section{Exciton Hamiltonian containing finite lifetime for the valence band hole}
In this section we derive the expression for the matter Hamiltonian (6) used in the article.

With the assumption of a frequency-independent final state Coulomb interaction, from the ladder diagrams in Fig. 2, a Bethe-Salpeter equation for the electron-hole bubble $P\left( {\bf k},i\omega \right)$ can be found along the lines of Ref. \cite{mahanBook}.
This integral equation for the vertex factor $P\left( {\bf k},i\omega \right)$   is given by
\begin{eqnarray}
F^{-1}%
\left( {\bf k},i\omega \right) P \left( {\bf k},i\omega \right)-\frac{1}{S}\sum_{{\bf k}_{1}}V_{{\bf k-k_{1}}}P%
\left( {\bf k}_{1},i\omega \right)  &=&g.
\label{eq:BSE}
\end{eqnarray}
The solution of this equation yields the photon polarization by the following relation
\begin{eqnarray}
\Pi(i\omega) &=& \frac{g}{S}\sum_{\bf{k}} P({\bf{k}},i\omega),
\end{eqnarray}
where for simplicity we have set the photon momentum equal to zero.
The function $F\left( {\bf k},i\omega \right)$ is given by the Matsubara summation over the internal fermionic frequency $i\nu_n$ of the electron and hole propagators, defined in Fig. 2. Because we consider a zero photon momentum, only electrons and holes with opposite momentum contribute. Hence we have
\begin{eqnarray}
F\left( {\bf k},i\omega \right) = -\frac{1}{\beta}\sum_{i\nu_n}G_e\left({\bf k},i\omega-i\nu_n\right)G_h\left(-{\bf k},i\nu_n\right).
\label{eq:F}
\end{eqnarray}
For the electron we use the free Green's function
\begin{eqnarray}
G_e({\bf{k}},i\nu_n)&=&\frac{1}{i\nu_n-\xi^e_{\bf k}},
\label{eq:Ge}
\end{eqnarray}
with $\xi^e_{\bf k}=\varepsilon^e_{\bf k} + \Sigma_{x}({\bf{k}})-\mu_e$. Here, $\varepsilon^e_{\bf k}=\hbar^2k^2/2m_e$ is the single particle dispersion in the parabolic mass approximation and $\Sigma_{x}({\bf{k}})$ stands for the screened exchange interaction of the optically excited electron with the Fermi sea. Also the electron chemical potential $\mu_e$ is renormalized due to exchange interaction.

For the hole Green's function, we include the interactions with the electron Fermi sea perturbatively as follows:
\begin{eqnarray}
G_h({\bf{k}},i\nu_n)  &=&  G^0_h({\bf{k}},i\nu_n)+ G^0_h({\bf{k}},i\nu_n)\,\Sigma^{RPA}({\bf{k}},i\nu_n)\,G^0_h({\bf{k}},i\nu_n).
\end{eqnarray}
Here, for the dynamical self energy $\Sigma^{RPA}$, describing hole relaxation that is responsible for a finite polariton lifetime, we use the 2D RPA result. The free hole Green's function is given by
\begin{eqnarray}
G^0_h({\bf{k}},i\nu_n)=\frac{1}{i\nu_n-\xi^h_{\bf k}}.
\end{eqnarray}
The hole dispersion is also assumed to be quadratic, so $\xi^h_{\bf k}=\hbar^2k^2/2m_h-\mu_h$. No exchange contribution shows up because we consider a single hole ($\mu_h=0$). 

In order to be able to perform the Matsubara summation in eq. \eqref{eq:F} exactly, we further approximate the RPA result to lowest order in the Lindhard polarization (`1 particle-hole bubble'), so only emission of a single particle-hole pair is taken into account:
\begin{eqnarray}
\Sigma^{RPA}({\bf{k}},i\nu_n) &= &  \Sigma_{1 bubble}({\bf{k}},i\nu_n) + \mathcal{O}(2\, \textrm{bubbles}).
\end{eqnarray}
Using the 2D expression of the Lindhard polarization function, the dynamical self energy is given by
\begin{eqnarray}
\Sigma _{1\,bubble}^{h}(\mathbf{k},i\nu _{n}) &=&-\frac{ 2 }{%
S^2}\sum_{\bf{K,q}} \,\left(V_{\mathbf{q}}^{s}\right)^2\,\left[ n_{F}(\varepsilon^e_{{\bf K}})-n_{F}(\varepsilon^e_{{\bf K+q}})\right] \frac{1-n_{F}\left[ \xi _{\mathbf{k}-%
\mathbf{q}}^{h}\right] +n_{B}(\varepsilon^e_{{\bf K+q}}-\varepsilon^e_{{\bf K}})}{i\nu _{n}-\xi
_{\mathbf{k}-\mathbf{q}}^{h}-\varepsilon^e_{{\bf K+q}}+\varepsilon^e_{{\bf K}}}.
\label{eq:LindhardPol}
\end{eqnarray}
Here, the exchange correction to the electron energy was neglected to simplify the subsequent calculations. Furthermore, the double summation runs over the center-off-mass momentum and relative wavevector of the particle-hole pair, resp. ${\bf q, K}$.\newline
Combining the expressions \eqref{eq:F}-\eqref{eq:LindhardPol} we can do the Matsubara sum over $i\nu_n$. The result is given by
\begin{eqnarray}
F\left( {\bf k},i\omega \right)&=&S_0\left(1+\frac{S_1}{S_0}\right),
\label{eq:Fshort}
\end{eqnarray}
where we have
\begin{equation}
S_0\left({\bf k},i\omega\right)=\frac{1-n_{F}\left( \xi^e _{\mathbf{k}}\right) -n_{F}\left( \xi _{\mathbf{-k}}^{h}\right) }{i\omega -\xi _{%
\mathbf{-k}}^{h}-\xi^e _{\mathbf{k}}}
\label{eq:S0}
\end{equation}
and
\begin{eqnarray}
\frac{S_{1}\left({\bf k},i\omega\right)}{S_{0}\left({\bf k},i\omega\right)}&=&
\frac{2}{S^{2}}\sum_{\mathbf{q,K}}\left(V_{\mathbf{q}}^{s}\right)^2\,n_{F}\left( \varepsilon^e _{%
\mathbf{K+q}}\right) \left[ 1-n_{F}\left( \varepsilon^e _{\mathbf{K}}\right) \right] \crcr
&&\times \left\{ 
\begin{array}{c}
\frac{1}{1-n_{F}\left( \xi^e _{\mathbf{k}}%
\right) }\frac{1}{-\xi _{\mathbf{-k}}^{h}-\left( -i\omega +\xi^e _{%
\mathbf{k}} \right) }\frac{1}{-\xi _{%
\mathbf{-k}}^{h}-\left( -\xi _{\mathbf{-k+q}}^{h}+\varepsilon^e _{\mathbf{K+q}}-\varepsilon^e _{%
\mathbf{K}}\right) } \\ 
+\frac{1}{1-n_{F}\left( \xi^e _{\mathbf{k}}\right) }\frac{1}{\left[ -\xi _{\mathbf{-k}}^{h}-\left( -\xi _{%
\mathbf{-k+q}}^{h}+\varepsilon^e _{\mathbf{K+q}}-\varepsilon^e _{\mathbf{K}}\right) \right] ^{2}}
\\ 
-\frac{i\omega -\xi _{\mathbf{-k}}^{h}-\xi^e _{\mathbf{k}} }{1-n_{F}\left( \xi^e _{\mathbf{k}} \right) }\frac{n_{F}\left( -\xi _{\mathbf{-k+q}}^{h}+\varepsilon^e _{%
\mathbf{K+q}}-\varepsilon^e _{\mathbf{K}}\right) }{-\xi _{\mathbf{-k+q}}^{h}+\varepsilon^e _{%
\mathbf{K+q}}-\varepsilon^e _{\mathbf{K}}+i\omega -\xi^e _{\mathbf{k}} }\frac{1}{\left( -\xi _{\mathbf{-k+q}}^{h}+\varepsilon^e _{\mathbf{%
K+q}}-\varepsilon^e _{\mathbf{K}}+\xi _{\mathbf{-k}}^{h}\right) ^{2}}%
\end{array}%
\right\} 
\end{eqnarray}
Upon analytical continuation $i\omega\to\omega-E_g-\mu_e-\mu_h+i\eta^+$ only the last line yields an imaginary part and
thus the lifetime of the hole. The other terms correspond to the energy shift, but these are beyond the scope of this work. One has
\begin{eqnarray*}
\text{Im}\frac{S_{1}}{S_{0}}&=&\frac{2\pi }{S^{2}}\frac{\omega -\varepsilon_{\mathbf{%
-k}}^{h}-\varepsilon^e _{\mathbf{k}}-\Sigma_x({\bf k}) }{%
1-n_{F}\left( \xi^e _{\mathbf{k}} \right) }\crcr
&\times&\sum_{\mathbf{q,K}}\left(V_{\mathbf{q}}^{s}\right)^2\,n_{F}\left( \varepsilon^e _{\mathbf{K+q}%
}\right) \left[ 1-n_{F}\left( \varepsilon^e _{\mathbf{K}}\right) \right] \frac{%
n_{F}\left( -\varepsilon _{\mathbf{-k+q}}^{h}+\varepsilon^e _{\mathbf{K+q}%
}-\varepsilon^e _{\mathbf{K}}\right) }{\left( \varepsilon _{\mathbf{-k+q}%
}^{h}-\varepsilon^e _{\mathbf{K+q}}+\varepsilon^e _{\mathbf{K}}-\varepsilon _{%
\mathbf{-k}}^{h}\right) ^{2}}\delta \left[ \omega -\varepsilon _{\mathbf{-k+q%
}}^{h}+\varepsilon^e _{\mathbf{K+q}}-\varepsilon^e _{\mathbf{K}}-\varepsilon^e _{%
\mathbf{k}}-\Sigma _{x}\left( \mathbf{k}\right) \right], 
\end{eqnarray*}
where the delta-function expresses conservation of energy and momentum. This is in fact expression (7) used in the article. The 4-fold summation can be worked out further to a 2-dimensional integral which we than compute numerically. This leads to the following result (all energies expressed in exciton Rydbergs $R_X$):
\begin{equation}
\text{Im}\frac{S_{1}}{S_{0}}=\Xi({\bf k}) \; \frac{\Sigma_{hole}({\bf k})}{\omega-k^2-\Sigma_x({\bf k})}
\label{ImS1S0}
\end{equation}
where the Pauli blocking factor $\Xi({\bf k})$ is
\begin{equation}
\Xi({\bf k})=\Theta\left(k>k_F\right)\Theta\left(\omega >\frac{k^2}{1+m_{e}/m_{h}}+\Sigma_{x}({\bf k})/R_{X}\right).
\label{eq:theta}
\end{equation}
Here the hole self energy is 
\begin{equation}
\Sigma_{hole}({\bf k})=\frac{1}{\pi}\frac{m_e}{m_r}\int_0^\infty dq\int_0^\pi d\theta\frac{\sqrt{k_F^2-a-K_{\text{min}}^2}-\sqrt{{\textrm {max}}(K_{\text{min}},k_F)^2-K^2_{\text{min}}}}{\left[q+s(q)\right]^2}
\Theta\left(a<0\right)\Theta\left(k_F^2-a>K^2_{\text{min}})\right)
\label{eq:SigmaHole}
\end{equation}
whith $s(q)$ the inverse screening length as defined in \cite{Stern}.
The functions $a({\bf k},q,\theta),K_{\text{min}}({\bf k},q,\theta)$ are given by
\begin{eqnarray}
a({\bf k},q,\theta)&=&\frac{m_e}{m_h}q^2 - 2\frac{m_e}{m_h}kq\cos\theta + \frac{m_e}{m_r}\left(k^2-\omega+\Sigma_x({\bf k})\right)\crcr
K_{\text {min}}({\bf k},q,\theta)&=&\frac{1}{2q}\left|a({\bf k},q,\theta)-q^2\right|
\end{eqnarray}
For the BSE the inverse of the function $F({\bf k},i\omega)$ was needed, see eq. \eqref{eq:BSE}. Combining eqs. \eqref{eq:Fshort}, \eqref{eq:S0} and \eqref{ImS1S0} we find 
\begin{eqnarray}
F^{-1}({\bf k},\omega) &=& S^{-1}_0\left(1+i\,{\text Im}\frac{S_1}{S_0}\right)^{-1}\crcr
&=&\left(\omega-k^2-\Sigma_x({\bf k})\right)\left(1+i\, \Xi\,({\bf k}) \frac{\Sigma_{hole}({\bf k})}{\omega-k^2-\Sigma_x({\bf k})}\right)^{-1}.
\end{eqnarray}
For a small hole self energy, we can use the following Taylor expansion: $1/(1+x)\approx 1-x$ for $x<<1$. Identifying $x=i\, \Xi\,({\bf k})\,\Sigma_{hole}({\bf k})/\left(\omega-k^2-\Sigma_x({\bf k})\right)$ we obtain Fermi's golden rule:
\begin{eqnarray}
F^{-1}({\bf k},\omega) &=&\left(\omega-k^2-\Sigma_x({\bf k})\right)\left(1-i
\,\Xi\,({\bf k})\,\frac{\Sigma_{hole}({\bf k})}{\omega-k^2-\Sigma_x({\bf k})}\right)\crcr
&=&\omega-k^2-\Sigma_x({\bf k})-i\,\Xi\,({\bf k})\,\Sigma_{hole}({\bf k}).
\end{eqnarray}
Plugging this expression into. eq. \eqref{eq:BSE} the identification with eq. (6) from the article is straightforward. It is worth noting that the Taylor series is consistent with the expansion of the RPA self energy to lowest order in the Lindhard polarization. Moreover, the polariton spectrum in figure 4 from the article shows that the linewidths are much smaller than the exciton binding energy. Hence, the current results and approximations are selfconsistent.

\section{Influence of final state interactions and Fermi edge}

\begin{figure}[hbtp]
\centering
\includegraphics[scale=0.5]{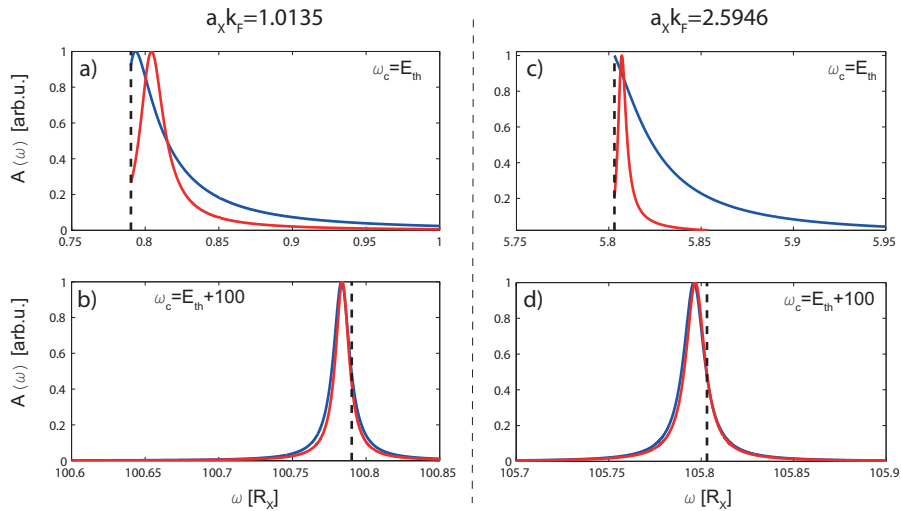}
\caption{Comparison of photon spectral functions with (red) and without final state interactions (blue) for different Fermi wave vectors. In order to appreciate the importance of the FSI the red and blue curves in figures a-c are to be compared. There a significant narrowing of the linewidth can be observed when FSI are taken into account. The role of the Fermi edge for the formation of the upper Fermi edge polariton is clear from the comparison of a-b and c-d. The `Rabi splitting' vanishes for a highly blue detuned cavity mode. Both the FSI and the Fermi edge are necessary for the formation of the upper polariton.} 
\label{fig:suppl}
\end{figure}

Both final state interactions and the presence of the Fermi edge are necessary for the formation of the upper Fermi edge polariton. This is illustrated quantitatively in Fig. \ref{fig:suppl} (a-b) and (c-d) for two electron densities and cavity detunings. The blue lines indicate the single-particle results while the red lines contain the final state Coulomb interactions.

Figures a,c show the photon spectral function on resonance with the Fermi edge $E_{\textrm {th}}$ (black dashed line) for two different electron gas densities. A clear suppression of the spectral function linewidth when taking into account final state interactions (FSI) is observed as compared to the single-particle result.

To estimate the importance of the Fermi edge, the cavity mode is largely blue detuned from the absorption threshold energy to $E_{\text {th}}+100$ (black dashed lines), see figures b, d. It is clear that both the single particle theory and final state interaction results now coincide (same scales are used to facilitate comparison). The Fermi edge is thus responsible for the `Rabi splitting', meaning that it really pushes the upper polariton into the electron-hole continuum. 

As a last remark, for very large blue detuning of the cavity mode from the Fermi edge, a red shift from the cavity mode is seen in figures b, d. This can be attributed to a different index of refraction in the electron-hole continuum in comparison with the excitonic region (see also Ref. \cite{Bajoni}). 

\end{document}